# A New Look at Type-III Bursts and their Use as Coronal Diagnostics

Samuel D. Tun Beltran[1] • S. Cutchin[2] • S. White[3]

[1]Naval Research Laboratory, Washington D.C. 20375, USA
email: Samuel.TunBeltran@nrl.navy.mil

[2]former National Research Council Fellow at the Naval Research Laboratory, Washington, DC 20375, USA

[3]Space Vehicles Division, Air Force Research Laboratory, Albuquerque, NM, USA

## ABSTRACT

We present meter-wave solar radio spectra of the highest spectro-temporal resolution achieved to date. The observations, obtained with the first station of the Long Wavelength Array (LWA1), show unprecedented detail of solar emissions across a wide bandwidth during a Type-III/IIIb storm. Our flux calibration demonstrates that the LWA1 can detect Type-III bursts much weaker than 1 SFU, much lower than previous observations, and that the distribution of fluxes in these bursts **varies with frequency**. The high sensitivity and low noise in the data provide **strong** constraints to models of this type of plasma emission, **providing evidence against the idea that Type-IIIb striae are generated from electrons trapped in Langmuir wave sidebands**. The continuous generation of electron beams in the corona revealed by the high density Type-III storm is evidence for ubiquitous magnetic reconnection in the lower corona. Such an abundance of reconnection events not only contributes to the total coronal energy budget, but also provides an engine by which to form the populations of seed particles responsible for proton-rich solar energetic particle events. **An active region (AR) with such levels of reconnection and the accompanying Type-III/IIIb storms is here proposed to be associated with an increase of SEP production if a CME erupts. The data's constraints on existing theories of Type-IIIb production are used to make an association of the observed Type-IIIb storm to specific electron beam paths with increased inhomogeneities in density, temperature, and/or turbulence. This scenario ties in the observed timing of Type-III and IIIb storms, constrained theories of Type-III and IIIb emission, and the ability of the emitting AR to produce a strong SEP event. The result requires but a single observable to cement these ideas, the statistical correlation of Type-III/IIIb activity with SEP-productive AR.**

**Keywords**: Corona, Radio Emission; Radio Bursts, Dynamic Spectrum; Radio Bursts, Type III; Waves, Plasma

## 1. Introduction

Type-III solar radio bursts are a common feature of low-frequency (< 300 MHz) solar radio emission during periods of solar activity (*e.g.* Wild and McCready, 1950; Kundu, 1965; Suzuki and Dulk, 1985). They are characterized by short durations (seconds) and a rapid drift from high to low frequency: in frequency–time plots of solar radio emission commonly known as dynamic spectra, they appear as intense nearly vertical features whose duration increase slightly at lower frequencies. Wild (1950) suggested that the frequency drift was due to the rapid outwards movement of a disturbance through the density and magnetic-field gradients in the solar atmosphere, and electron beams were believed to be the most likely source. Ginzburg and Zheleznyakov (1958) developed the theory for emission of electron beams in a plasma via the generation of electrostatic Langmuir waves at the local plasma frequency and their conversion to propagating electromagnetic waves at the fundamental and harmonic of the plasma frequency. Since the plasma frequency is proportional to the square root of the ambient electron density, outwards motion of an electron beam would naturally lead to a frequency drift. Subsequently direct *in-situ* observations of electron beams by satellites in the solar wind confirmed that they are the source of Type-III radio bursts (*e.g.* Lin, 1985).

Type-III bursts are often seen at the onset of impulsive solar flares, but also occur during periods of high solar activity outside flare times. So-called "Type-III storms" can last hours or days and consist of almost continuous Type-III activity. When observed at high frequency and time resolution, Type-III bursts can also show significant structure. Both fundamental and harmonic traces can be seen on dynamic spectra, but in addition there are bursts whose overall envelopes track the frequency drift typical of Type-III bursts but the emission within the envelopes consists of short, narrow-band features intermittent in both time and frequency. These small bursts are known as "striae", and a burst composed of such features is known as a Type-IIIb burst (de la Noe and Boischot, 1972).

Here we present preliminary work that demonstrates the remarkable value of the high temporal and spectral resolution of the LWA1 station for solar studies, with impact on derived physical parameters. During the LWA1 observing campaign number LT003 we captured a storm of weak Type-III bursts. The LWA1 sensitivity is such that it captured many details not seen by other radio observatories, such as the Solar Radio Spectrograph (SRS) at Sagamore Hill, Massacjisetts. USA, one of the standard databases from which solar activity is reported to the community. We present the greatly detailed observations along with preliminary analysis showing the power of this new array for the study of the solar corona.

## 2. Observations and Flux Calibration
### 2a. Observations

The Sun was observed with the first station of the Long Wavelength Array (LWA1) on 12 – 15 April 2013. At the time of observations the LWA1 station consisted of 256 quadrupole antennas over an area of side dimension 100 m and was in its commissioning phase. The spatial resolution of a single LWA station cannot resolve the radio Sun, so analysis will focus on the spectro–temporal properties of solar radio emission. Work done to characterize the LWA1 station demonstrated that the system equivalent flux density of the array was approximately 6 kJy at Zenith and 8 – 10 kJy at 15 degrees from the Zenith for frequencies between 30 and 74 MHz (Schinzel and Polisensky, 2014). We therefore observed the Sun for four hours around local noon (16 – 20 UT) to optimize the sensitivity of the array. We observed in the digital receiver spectrometer (DRS) mode by beam-forming four 19 MHz beams, using the instrument's gain setting of six. Two beams with complementary frequency coverage pointed in the direction of the Sun in order to give full 10 – 75 MHz band coverage and two others pointed at the Zenith to obtain data for radio frequency interference (RFI) anti-coincidence. The instrument recorded data at 100 ms temporal, and 4.9 kHz spectral resolutions. A flux-calibrated bandpass correction was then applied to the data, the details of which are in the next section.

For comparison we also use daily observations of the Sun from the US Air Force's Solar Radio Spectrometer (SRS) at Sagamore Hill, Massachusetts. The frequency range matching that of the LWA1 is observed with the SRS's Semi-Bicone antenna, and the data products are total power in the 25 – 75 MHz range, with a three-second temporal and 125 kHz frequency resolution. A log-periodic antenna records the 75 – 180 MHz range (Air Force Weather Observer, 2011). A bandpass calibration has been applied to this data, and all plots will show the optimum contrast of low-flux features. The overlapping coverage of the LWA1 and SRS data is from 25 to 77 MHz. The Sun is also observed daily by the Decameter Array (DAM) at Nançay, France. This array consists of 144 conical helicoil antennas detecting right and left cicular polarizations from 10 – 100 MHz. Solar data are usually examined from 25 – 80 MHz. Alternating right and left polarizations are recorded every 494 ms with a 138 kHz frequency resolution. Because of several issues with RFI mitigation and the implementation of bandpass normalization, we only use DAM data to explore the occurrence of Type-III activity, but not for quantitative work. **Table 1 compares the basic instrument parameters for these instruments, highlighting the increased capabilities of the LWA1 station.**

| Observatory | Frequency Range [MHz] | Frequency Resolution [kHz] | Temporal Resolution [sec] | # of antennas |
|---|---|---|---|---|
| Sagamore Hill | 25 - 75 | 125 | 3 | 1 |
| Decameter Array | 10 - 100 | 138 | 0.494 | 144 |
| LWA1 | 10 - 75 | 4.9 | 0.1 | 258 |

**Table 1. Comparison of basic instrument parameters for the three observatories used in this study.**

The LWA1 detected a long-duration Type-III storm on all four days of the April 2013 observations. A comparison of LWA1 and SRS data makes it clear that this forest of Type-III is largely undetected by the SRS instrument, with only the occasional stronger individual Type-III burst being evident. DAM quick-look data clearly show that this Type-III storm is in every day's observation starting from 31 March 2013. 14 April is the last day of the Type-III storm as observed by the DAM array, but the LWA1 sees a continuation of the low-flux Type-III storm into 15 April when a type I storm also appears. On 14 April the Type-III storm is enhanced by a Type-IIIb storm. Despite a less clear detection of the Type-IIIb bursts, the SRS data shows that the Type-IIIb storm begins around 10:30 UT. Since this Type-IIIb emission continues until the end of both storms, we expect that the onset of the Type-IIIb bursts should be accompanied by a permanent change to the conditions required to produce them. No such changes are readily evident in the activity leading up to this time. AR 11718 produced multiple daily C-class flares from 6 – 14 April and single C-class flares on 15 – 16 April, but no major changes to magnetic or coronal structures are observed. On 11 April at 6:55 UT AR 11719 produced a large M6.5 flare and a coronal mass ejection (CME) which was accompanied by a strong SEP proton event. However, the prolonged period from these events and the onset of the Type-IIIb storm makes a direct relationship between them appear unlikely. Still, there will be reason to return to the possible role of these later events on the Type-IIIb emission in the discussion section of this article.

## 2b. Flux Calibration

A flux-calibrated bandpass normalization is needed to take care of most of the variations in the instrumental response, and it will convert the instrument's voltage data into the received flux density from solar emissions. We carry out a flux calibration of our solar data using LWA1 Crab Nebula observations taken with the same gain settings (here gain=six). For the Crab Nebula observations, we first generate a lower-envelope spectrum using the median of the lowest 200 points in each frequency channel to create a lower-envelope (LE) spectrum. This LE spectrum is used because strong solar emissions may still make it into the LWA1 beam pointing at the Crab. We then filter this further to remove narrow RFI features to produce a smooth response. The nominal Crab nebula source spectrum (a power law with spectral index 1.75; Bridle, 1970)is then folded by the smoothed response to get the flux-calibrated bandpass correction. There is a discontinuity in the BP correction around 55 MHz (Figure 1). This discontinuity is appropriate for this data set because there is an equal variation in the response in those bands.

We next find the correction factor of the solar data by matching its own inverted LE spectrum to the flux calibration determined above. For this we use the solar data of 13 April 2013, a Type-III storm day with no other major activity. Considering the dip around 34 – 35 MHz common in both spectra we find that a factor of 800 applied to the solar data brings this feature in both spectra into agreement (Figure 1). Considering the lack of a discontinuity of the solar LE spectrum, the same calibration factor should be applicable to all of the bandpasses. Since all of these solar observations were made with equal setting, this single correction factor is then used to bootstrap the flux calibration to all the solar observations. At the lower frequencies, the solar data are affected by large amounts of RFI noise throughout, so we are generally ignoring the data below 30 MHz other than for qualitative purposes, **as only the lowest bandpass is greatly contaminated by such interference (Figure 1, below 28 MHz). The plot also shows there is no discernible amount of RFI sources above 30 MHz in the LWA1 data at this time, and we find this to be true throughout the observations.**

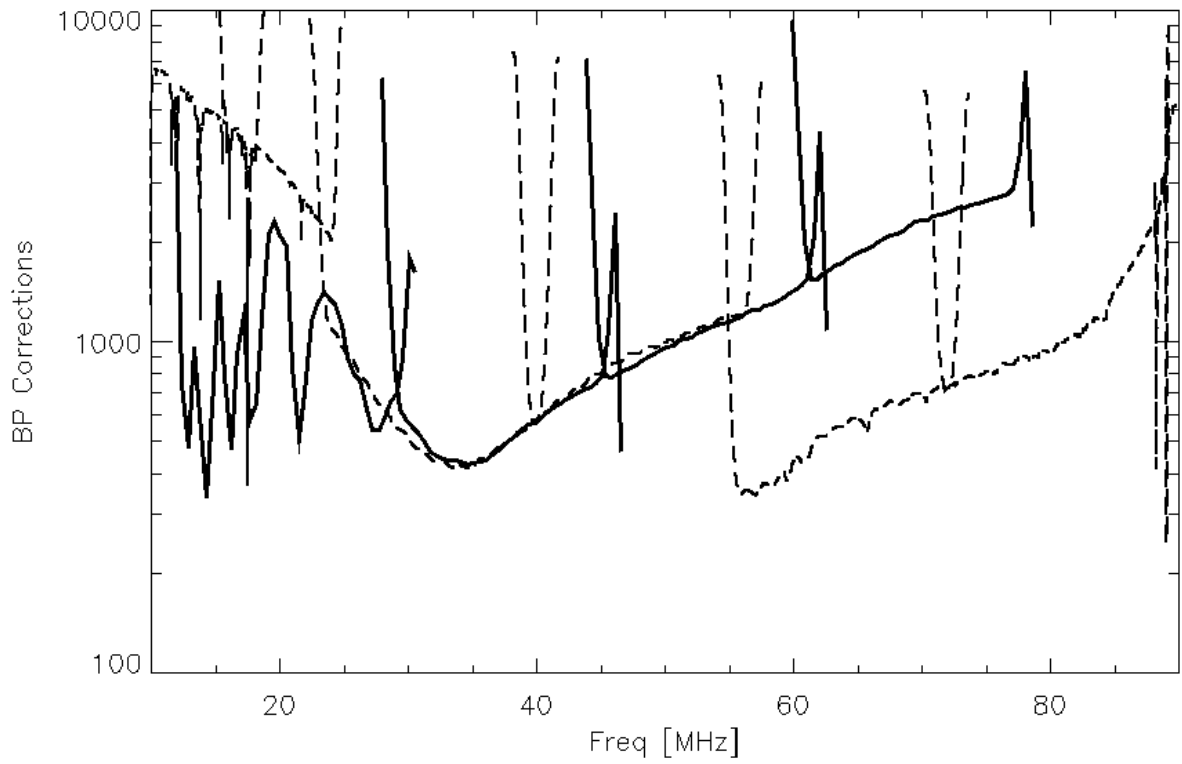

Figure 1. The low-pass envelope of solar observations (four bandpasses in solid dark lines), and the flux-calibrated low-band pass observations of the Crab Nebula (dashed lines). A multiplicative factor brings the features from 35 – 50 MHz in both bandpasses into agreement.

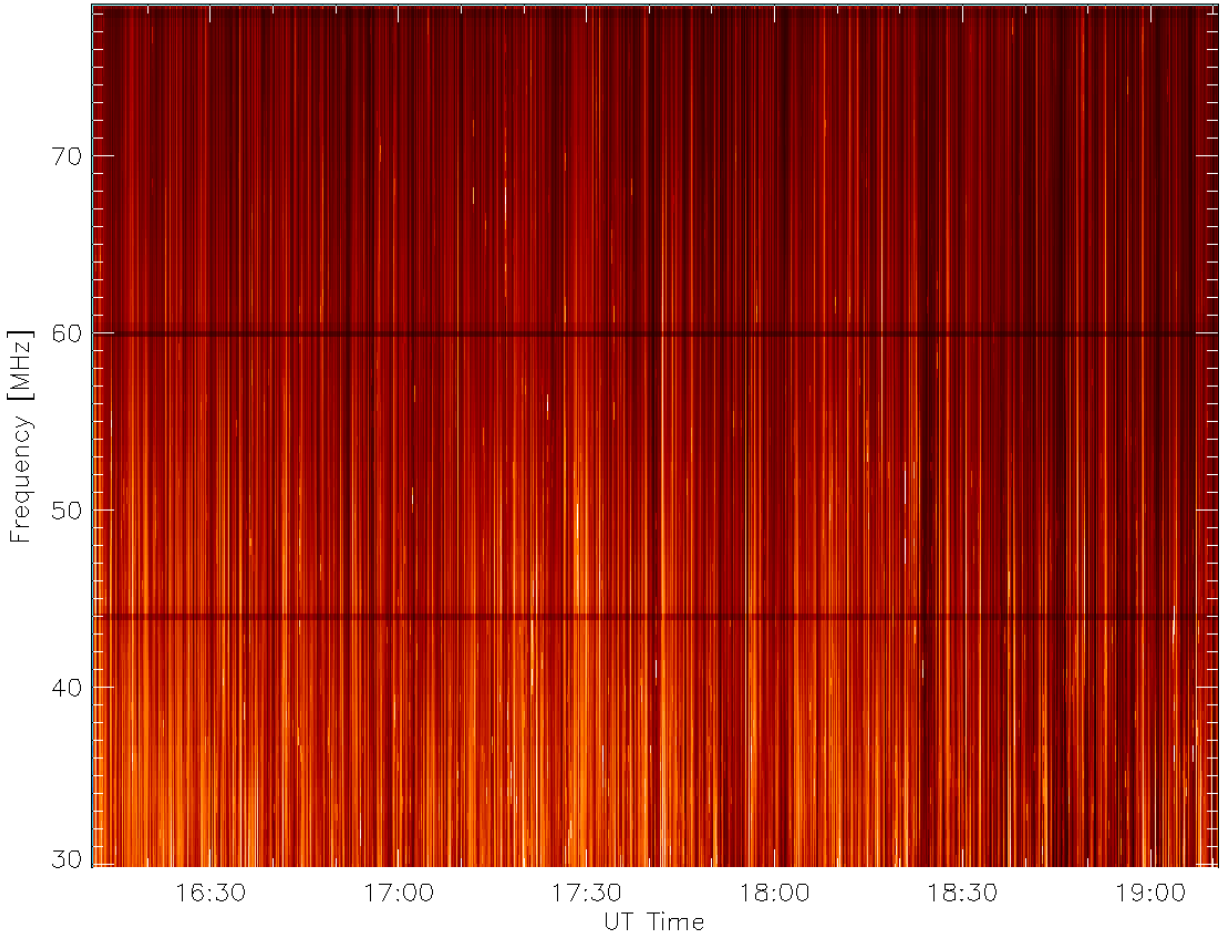

Figure 2. The flux-calibrated LWA1 dynamic spectrum for three hours during the 13 April 2013 observations. The vertical features throughout are Type-III bursts.

### 3. Analysis

### 3a. Type-III storm

The outstanding sensitivity of the LWA1 data permits us to analyze the Type-III/IIIb emissions at high temporal and spectral resolution. In order to distinguish the Type-III and the strongest IIIb activity, we characterize the Type-III storm with the data from 12 – 13 April, before the onset of the Type-IIIb storm. The data is integrated to 0.1 second temporal and 0.48 MHz frequency resolution to facilitate the automation of a burst-detection algorithm. In addition, a double polynomial smoothing over one second was performed for the same purposes. The data around the returned peaks were fitted with a Gaussian form. Only those peaks that were surrounded by Gaussians with width > 0.3 seconds, flux > 0.3 SFU, and a minimum separation between bursts of 0.5 seconds were retained. These experimentally determined parameters allow the detection of most local peaks corresponding to Type-III bursts while minimizing the inclusion of Type-IIIb (short duration and bandwidth) emission. Some Type-IIIb bursts are still included, and some very small Type-III bursts are excluded, so we estimate the relative error through the variability in the total number of bursts detected per frequency interval. We find that the maximum variation below 71 MHz is 4.4 %, and the counts for 75 MHz are off by 14.2%, with most of the errors resulting from the emission at lower fluxes. The resulting burst distributions for 13 April are shown in the dynamic spectrum in Figure 2. The storm produces Type-III bursts at a rate of ≈800 bursts per hour. The vast majority of these burst have fluxes <= 1 SFU, but it is important to note that the flux of a given burst increases with decreasing frequency. This causes the distribution of bursts with respect to peak flux to vary with frequency as well. Figure 3a shows a small segment of spectral data with various colored lines depicting cuts across various frequencies. Figure 3b shows the histogram of burst number vs. flux for these frequencies during the entire observation on 13 April, with the same colors as in Figure 3a. We have already demonstrated that our calibration method flattens the LWA1 bandpass, and the number of bursts does not increase with diminishing frequency. Figure 4 shows the low-level emission in the solar observations, the flattened and calibrated low-level bandpass, and the median of the entire day's observation. It shows that the bandpass was properly taken into account and that the rise in emission toward lower frequencies is not an instrumental effect. This measurement is consistent with previous results, which found an increase in flux from the Type-III bursts as they travel towards lower frequencies (greater heights above the Sun's surface; *e.g.* Suzuki and Dulk, 1985).

Noting that the flux-density distribution changes greatly with frequency, we compare the region more commonly studied in other works, where the distribution is best fit by a power law. This region is found to be common in all of the distributions from ≈ 10 – 100 bursts. An example of the power-law fit to the 75 MHz distribution can be seen in Figure 3b. We find that, although the errors are large, there appears to be a weak trend of diminishing power-law index towards lower frequencies (Figure 5). To study the frequency-dependent variations of individual bursts, we characterize the temporal profile for a single Type-III burst at several frequencies. Because of the large Type-III burst rate and scattered Type-IIIb activity it was difficult to find many such examples, so we concentrate on a single isolated burst for this analysis. Figure 6a shows our analyzed burst with some of the single-frequency cuts used, demarcated by dashed lines. Figure 6b shows the single frequency profiles for 73.3, 64.9, 60.6, 53.7, 48.9, 31.4, and 28.7 MHz at 0.01 second resolution and a 20 kHz bandwidth. Using the inflection point immediately after the initial linear rise in flux as a marker and assuming harmonic emission, the exciter travels from a density of $1.6 \times 10^7$ cm$^{-3}$ to $2.5 \times 10^6$ cm$^{-3}$ in 3.2 seconds. That is, the flux density received at Earth increases by a factor of three over a six-fold drop in the local plasma density at the emission site.

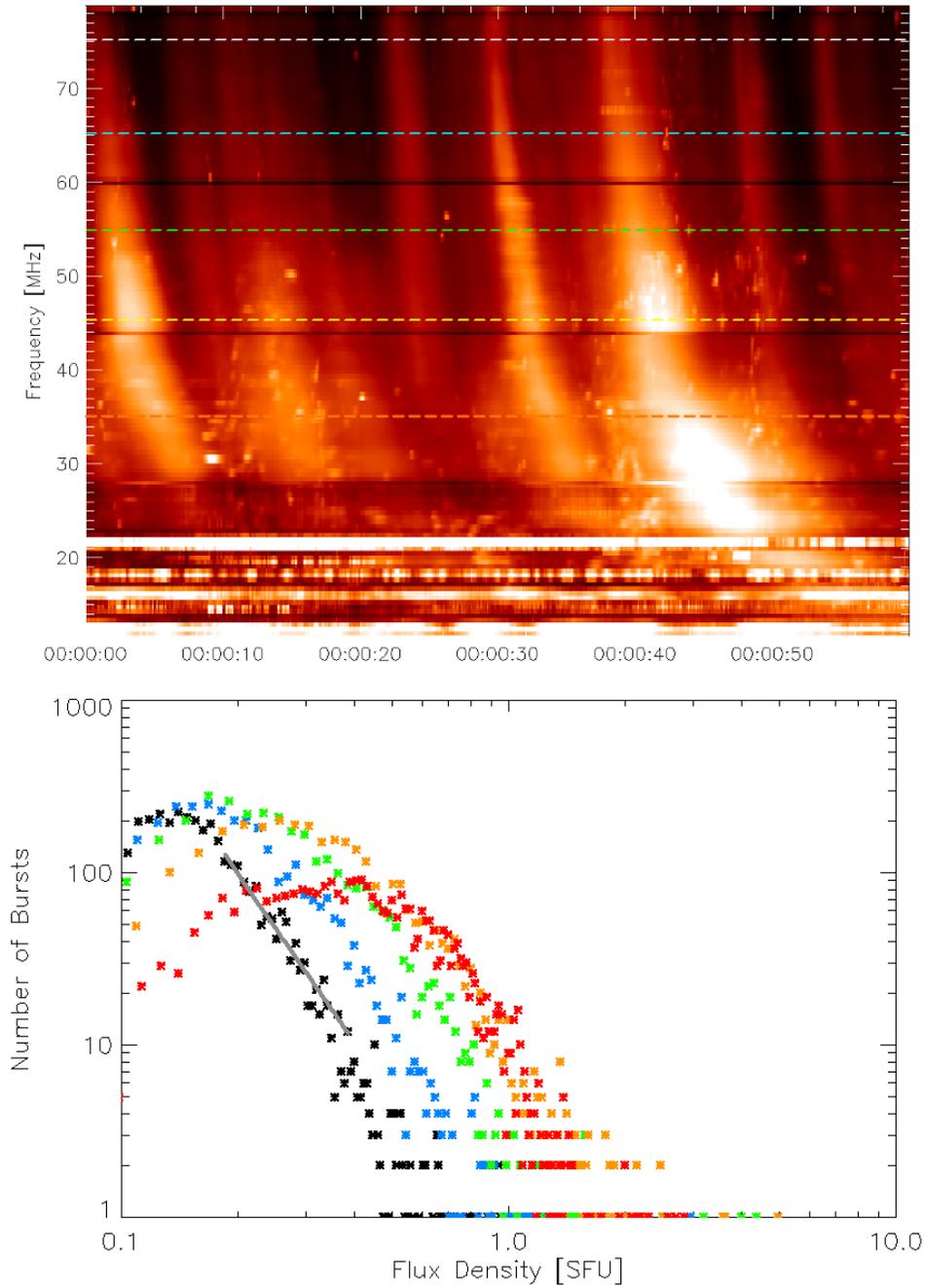

Figure 3 top) 60 second segment of data showing burst detail and the frequency cuts used in bottom) histogram of burst flux for the entire observation of 13 April, 2013. Black, blue, green, orange and red points corresponds to the frequencies 75, 65, 55, 45, and 35 MHz, respectively. The solid-gray line represents the power law fit to the 75 MHz distribution between ≈ 10 – 100 bursts.

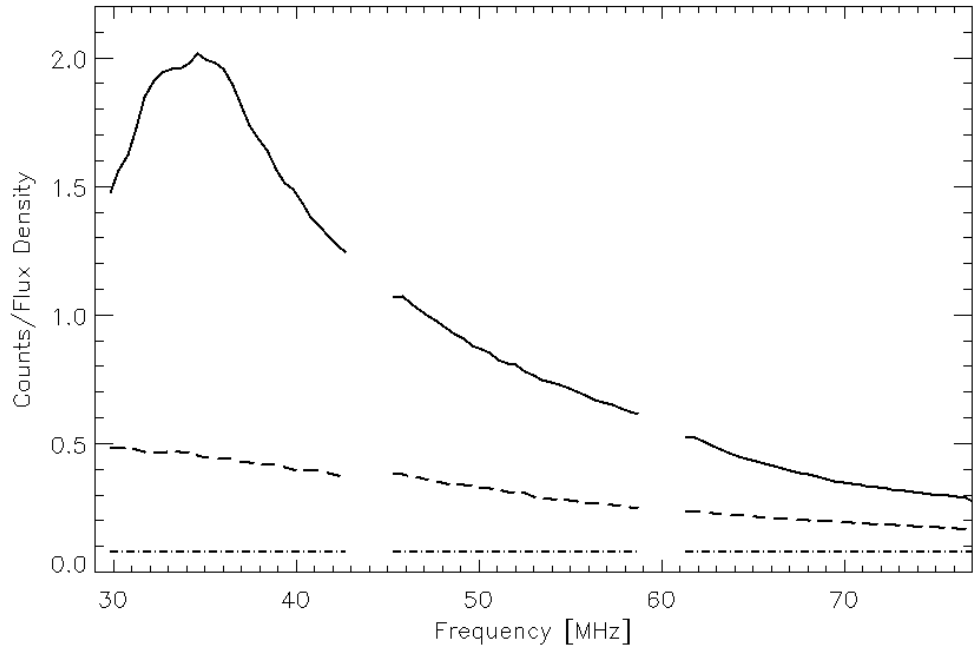

Figure 4. The median low level emission (solid), the flattened and calibrated bandpass (dashed), and the median of the total emission (dashed–dot) in the solar observations for 13 April 2013.

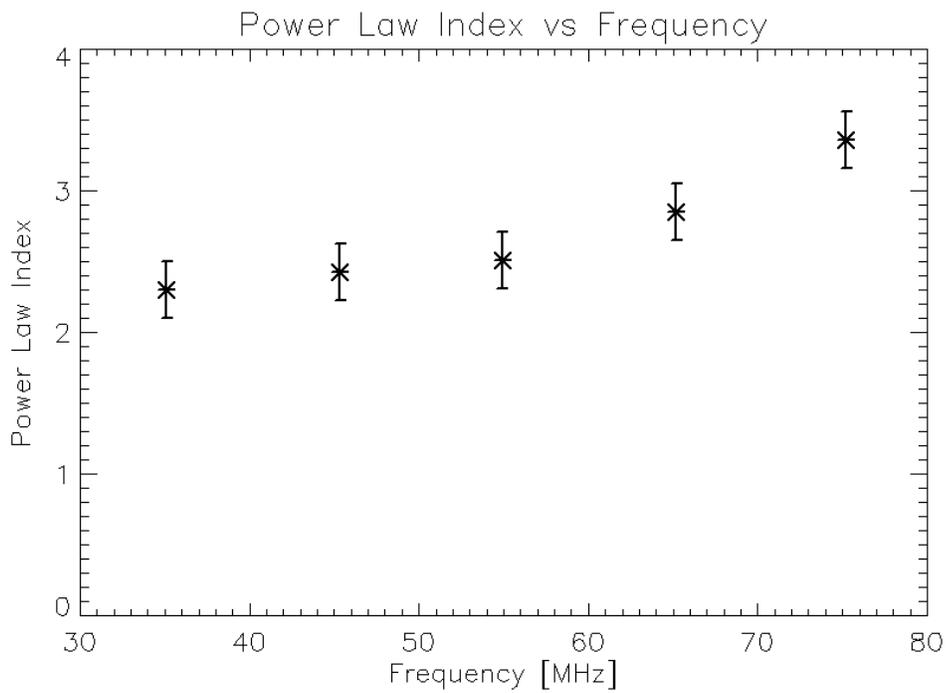

Figure 5. Power-law coefficient distribution vs. frequency.

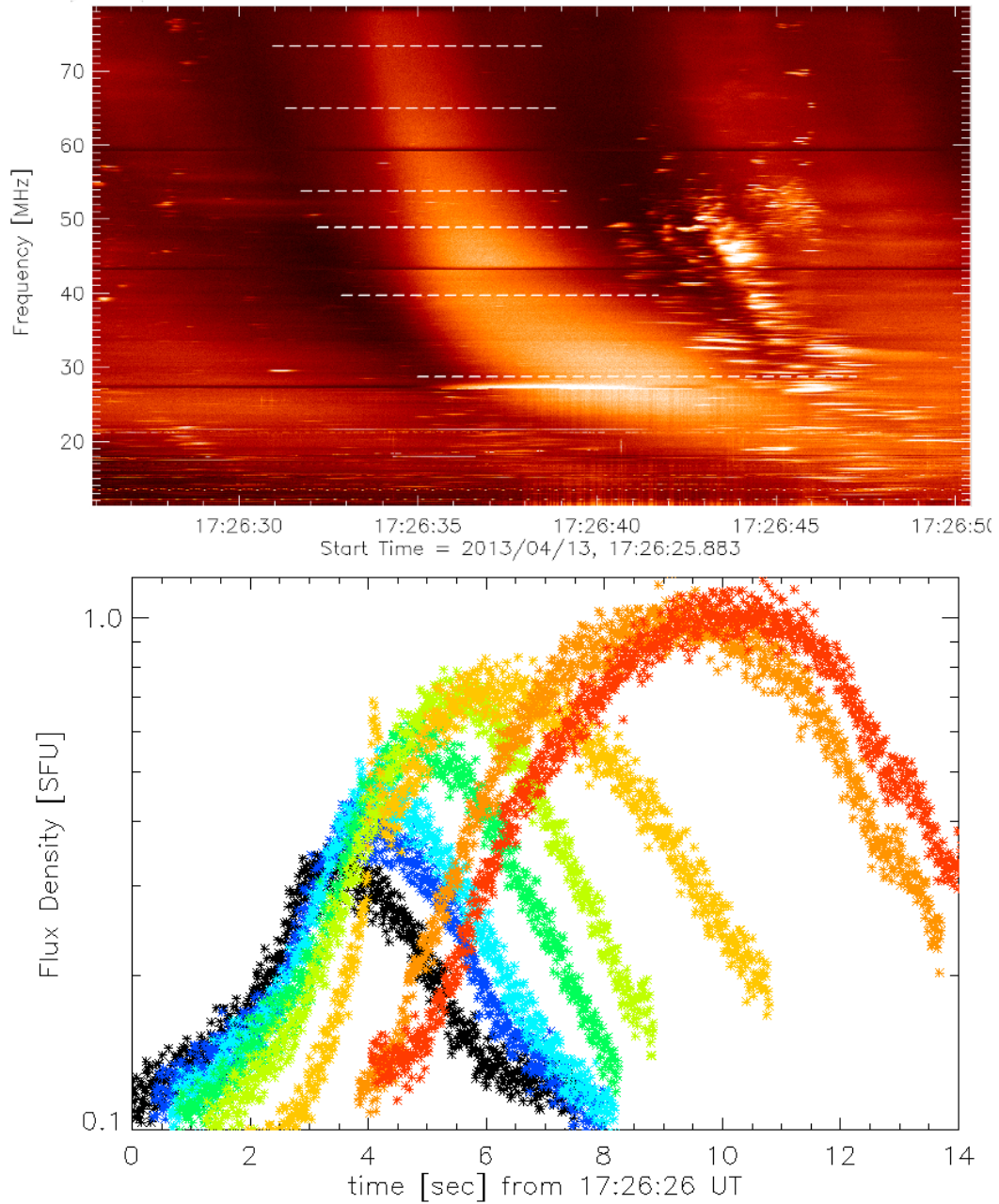

Figure 6 top) Isolated Type-III burst with frequency cuts represented by dashed lines, bottom) single-frequency profiles of the Type-III burst for (black to red) 73.3, 64.9, 60.6, 53.7, 48.9, 31.4, 28.7 MHz.

Noting that the initial rise in flux is exponential (linear in the log plot), we approximate a shift to the reference frame of the exciter, the traveling electron beam's bump in the tail distribution, by aligning this feature in the burst cuts for 73.3, 64.9, 60.6, 53.7, 48.9, and 39.6 MHz (Figure 7). This represents a drop in density from $1.6 \times 10^7$ cm$^{-3}$ to $7.4 \times 10^6$ cm$^{-3}$. The short peak in the 39.6 MHz profile at 1.75 seconds is from Type-IIIb emission and can be ignored for this analysis. The slope in the initial rise in flux density (zero – two seconds) steepens toward higher frequencies, as does the plasma emission damping phase (> four seconds) for 73.3 to 48.9 MHz. The slope of the damping phase is shallower for the 39.6 MHz profile. The steepening initial slopes could represent a more effective growth rate of the responsible Langmuir waves, a more efficient conversion of these into electromagnetic radiation, or a decrease in the opacity of the plasma at lower densities. The electron beam is not expected to gain energy as it travels outward in the corona, so the increased amplitude of the burst must be a property of the emission mechanism. This could be the efficiency of the exciter in driving Alfvén waves, the efficiency of the coalescence of waves to produce electromagnetic radiation, or a reduction in opacity, which allows more of the created radiation to escape. It is tempting to use the duration of the burst at each plasma level to determine the length of the exciter, but as Aubier and Boischot (1972) have pointed out, we will first need to deconvolve it from the plasma response. The LWA1 data are of sufficient quality to allow us to further explore the theoretical possibilities.

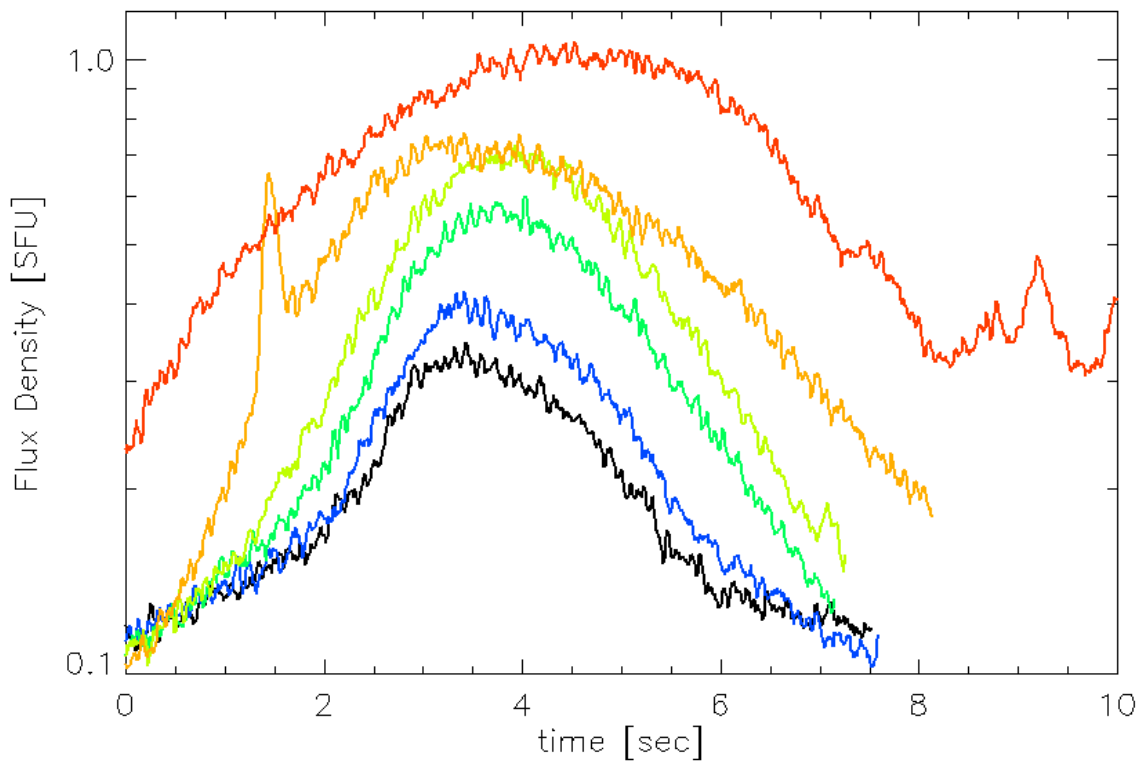

Figure 7. Single-frequency burst profiles for 73.3, 64.9, 60.6, 53.7, 48.9, and 39.6 MHz (black to orange, bottom to top) MHz, shifted to align the initial, linear rise in flux. This approximates the reference frame of the exciter. The red line depicts the 28.7 MHz emission shifted to an earlier time to demonstrate the larger envelope.

### 3b. Type-IIIb storm

On 14 April the LWA1 very clearly detected a highly polarized Type-IIIb storm in great detail (Figure 8). Observations from the previous days show occasional Type-IIIb bursts, but not in the number and rate suddenly appearing on 14 April (see Figure 2). Only the brighter bursts appear in the SRS data (Figure 9), with the DAM yielding a better indication of the ongoing storm (Figure 10). The Type-IIIb bursts are mostly confined to 60 – 18 MHz, and are strongly left-hand polarized. Figure 11 provides greater detail of the Type-IIIb bursts around the time a J-burst occurred, along with their polarization signal. J-bursts are thought to be emitted by the same mechanisms as Type-III bursts, but the electron beams that trigger them travel along closed field lines that revert their trajectory back towards the Sun. The J-burst occurring at 19:03:15 UT is completely unpolarized, while the surrounding Type-IIIb bursts are strongly left-hand polarized. Although dispersion and scattering effects have a great influence at these frequencies and densities, fundamental plasma emission is generally more polarized than at the second harmonic. Several studies have concluded that Type-IIIb bursts are fundamental emission phenomena that are sometimes accompanied by second-harmonic emission in the form of regular Type-III bursts (*e.g.* Ellis and McCulloch, 1967; Takakura and Yousef, 1975; Melnik et al., 2011). Type-III bursts at this time show slight to no polarization or clear harmonic relation.

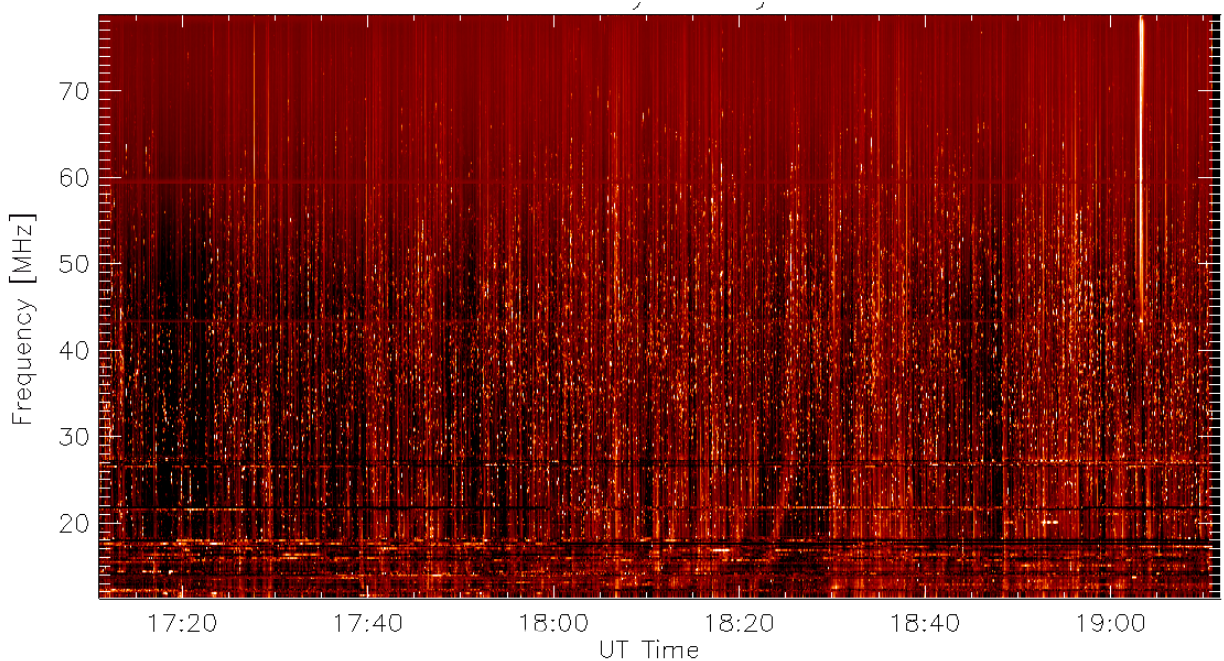

Figure 8. LWA1 left-handed polarization data for 14 April 2013. The sparkling appearance of theLWA1 data is due to the Type-IIIb storm, and stands in contrast to the spectra from 13 April 2014 in Figure 2.

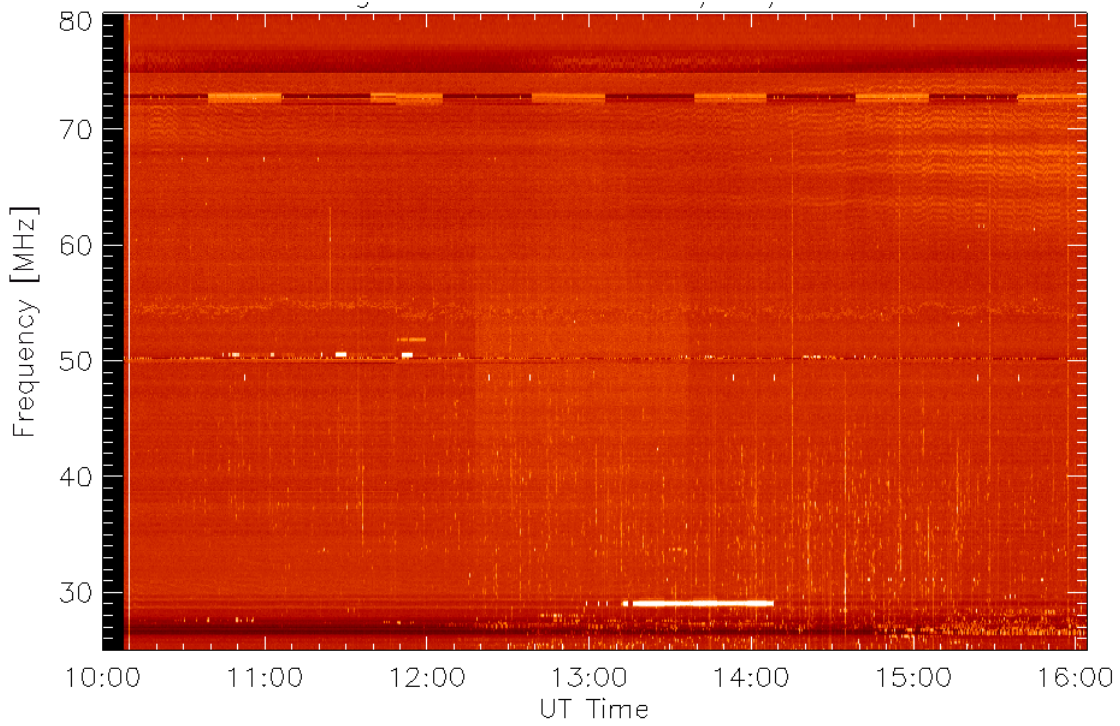

Figure 9. SRS total power dynamic spectrum for 14 April, 2013. The Type-III and -IIIb bursts are near the noise level of the instrument, so that only the brightest bursts are visible.

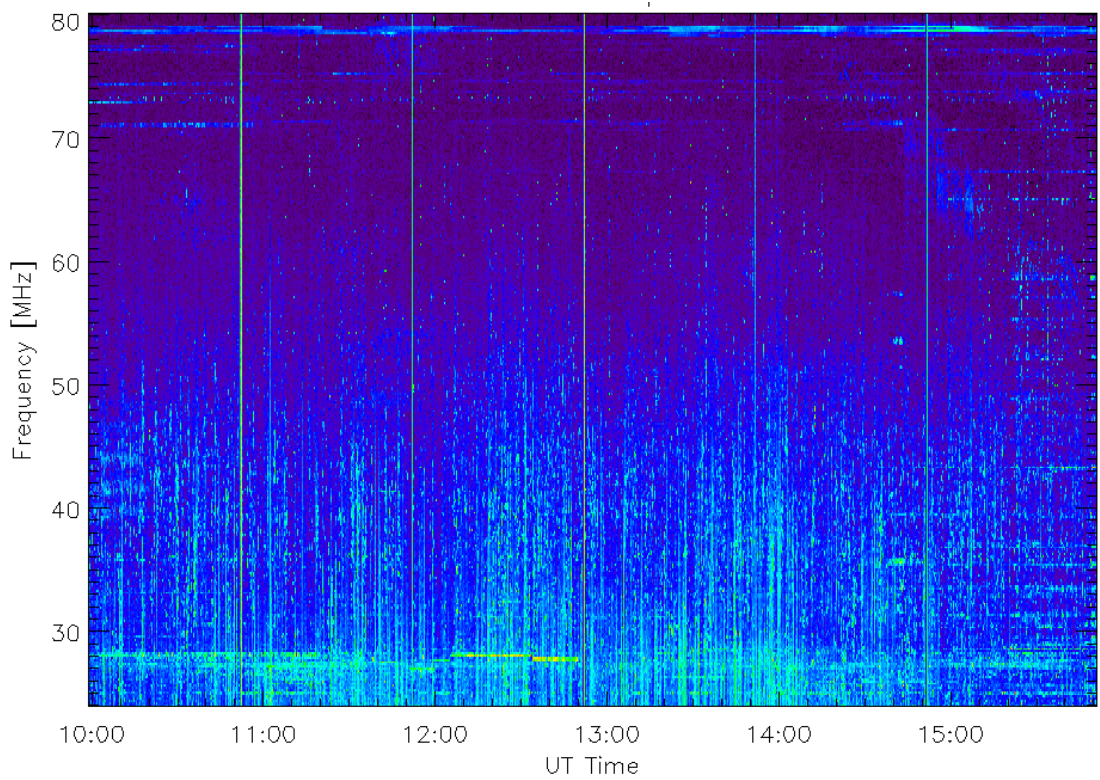

Figure 10. DAM left-handed polarization dynamic spectrum for 14 April, 2013.

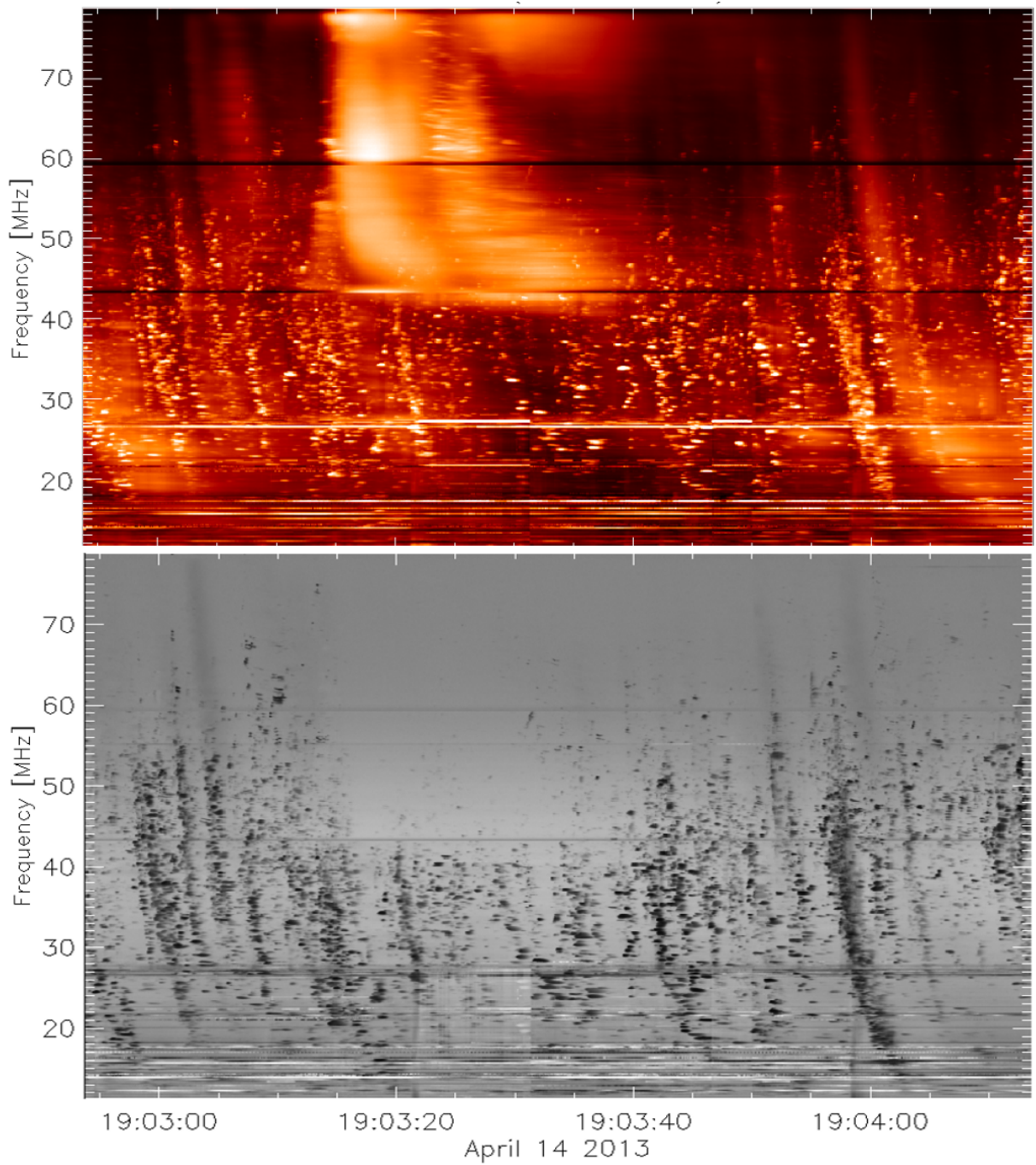

Figure 11. (a) LWA1 calibrated Stokes-*I* data around a J-burst , and (b) the corresponding degree of polarization spectrum on 14 April 2013 (black is left-handed polarization).

**3c. Type-IIIb Striae**

The resolving power and sensitivity of the LWA1 station is best demonstrated by the analysis of individual stria/striae components of Type-IIIb bursts. We do not have the highest resolution form of the data for the day of the Type-IIIb storm, 14 April. Still, there are sufficient Type-IIIb bursts in the other days to analyze. Figure 12 shows three Type-IIIb bursts on 13 April, along with a few regular Type-III bursts in the background and some sporadic striae groups below ≈ 40 MHz. The stria chain of the central Type-IIIb burst extends to lower frequencies than the other two. Figure 13 gives details of this central burst between 17:12:04 – 17:12:10 and 54 – 59 MHz. Clearly there is great variation in the amplitude, bandwidth, duration, separation of multiple components, and distribution of emission within this small segment of time. In addition to resolving many details of the stria emission, the good signal-to-noise in the dynamic spectra allows further analysis.

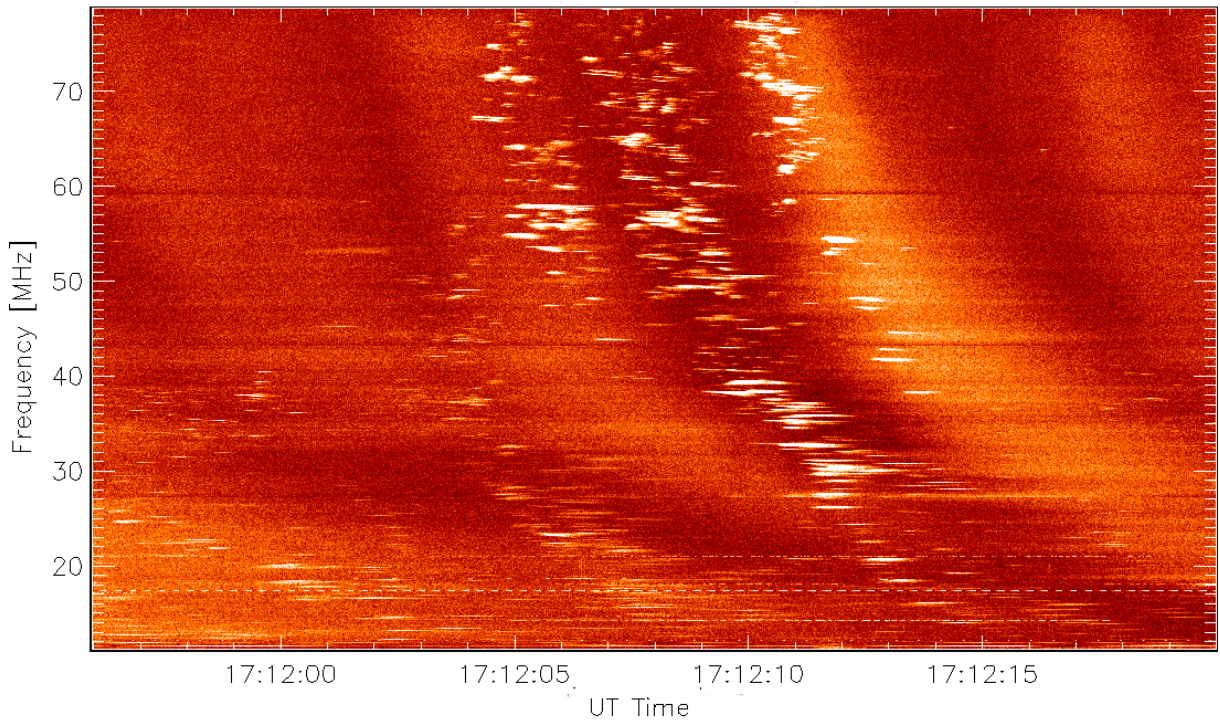

Figure 12. Example of a Type-IIIb bursts, Type-III bursts, and sporadic stria groups, shown here in left-handed circular polarization.

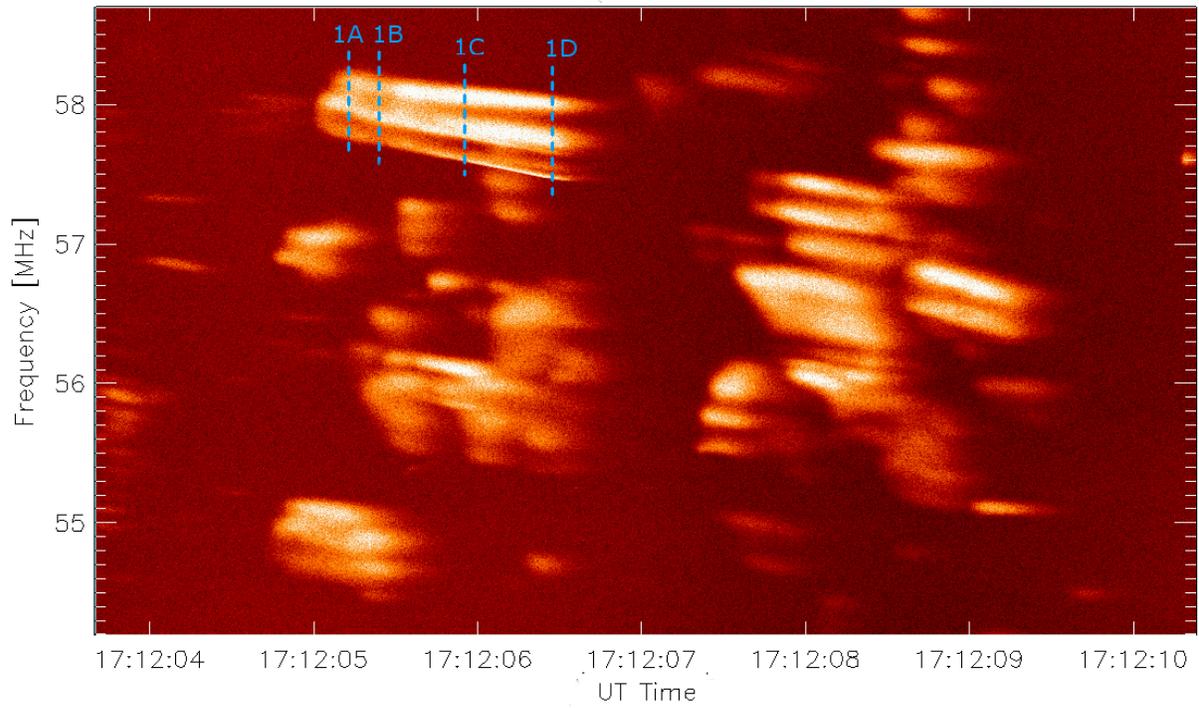

Figure13. Details of the central burst in Figure 11. There are single, double, and triple stria bursts of varying amplitude, duration, and spectral drift. The vertical lines denote the spectral cuts displayed in Figure 13.

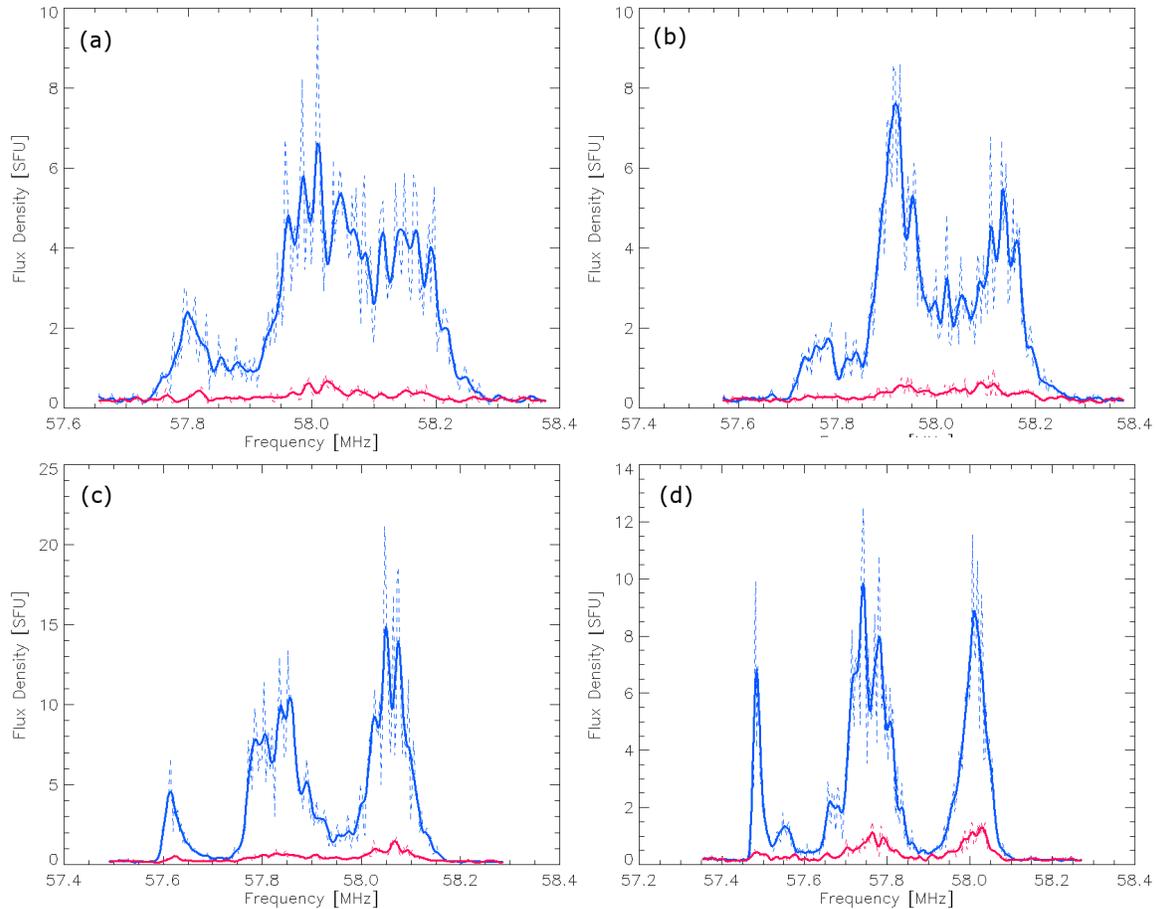

Figure 14. Spectral cuts across a single stria group, as marked upon Figure 12. Blue is left-handed and red is right-handed polarization. The flux density is in SFU.

Figure 14 shows the instantaneous spectral cuts corresponding to the vertical blue lines in Figure 13. The cuts span across a relatively long-lived triple burst that exhibits variations in amplitude and drift rate. The amplitudes of all three components increase with time, and at all times have a higher flux density than most of the Type-III bursts observed during these days. We note that the relative amplitude of the high and middle frequency striations (from the upper sideband and the main wave, respectively) invert in time. In fact, the relative amplitudes of these two highest components of all triple bursts in this Type-IIIb burst vary greatly and show no preference for the dominance of any either component. The lowest frequency striation, from the lower sideband, is observed to always have a lower flux density than both of the other two.

Smith and de la Noe (1976) proposed that oscillations of electrons trapped in the sidebands of beam-generated Langmuir waves could grow and coalesce, decaying into electromagnetic radiation in up to three frequency-separated sources, effectively producing striae emission. One of the kinematically derived conditions of their theory is that at all frequencies and at all beam velocities the opacities of the upper sideband, main wave, and lower sideband optical depths increase in that order ($\tau^+ < \tau^0 < \tau^-$). The intensity of plasma emission does not directly give any information about the energy of the responsible electrostatic waves, but rather of the length of the path along which the emission was allowed to grow. Solution of the equation of radiative transfer at these radio wavelengths, where the intensity is proportional to the brightness temperature, gives the relation

$T_b = T_o e^{-\tau} + T_{eff}(1 - e^{-\tau})$

(*e.g.* Dulk, 1985) where $T_b$ is the brightness temperature, $T_o$ is the effective temperature of the source, and $T_{eff}$ is the effective temperature of the intervening plasma. The proposed emission is from the main wave and its two sidebands, so that the emission from all three emanates from approximately the same physical volume. Since we expect that all three transverse waves will experience approximately the same intervening conditions along the path to the observer, the relative differences in the brightness temperatures observed should be mostly due to differences in the optical depths at the source. Further, the LWA1 beam sizes are larger than the source sizes, so the flux density received is proportional to the brightness temperature integrated over the small source sizes. Thus, the observed flux density from each striation in a triple should be in the order $\exp(-\tau^+) > \exp(-\tau^0) > \exp(-\tau^-)$. The profiles of the Type-IIIb striae in the LWA1 data discussed here may be evidence against this mechanism of Type-IIIb generation since the upper sideband and main wave emission do not show any preference for which striation exhibits the highest flux density.

Going down in frequency along this same Type-IIIb burst, an additional, different type of striae appear around ≈ 42.3MHz . Figure 15 shows details of two such stria groups. The striae are narrower in bandwidth than the examples discussed previously, and the frequency separation between components is also reduced. There is also more noticeable variation in frequency drift along each individual stria, imparting upon them the appearance of undulation. This wavy appearance might be explained if the exciter of the striation were to encounter density inhomogeneities along its path. Takakura and Yousef (1975) proposed such a scenario to explain the intermittency of emission along a Type-IIIb burst (the distribution of stria along a Type-III-like path). We propose it here not to explain the occurrence of stria groups but rather to explain the undulations along individual striations.

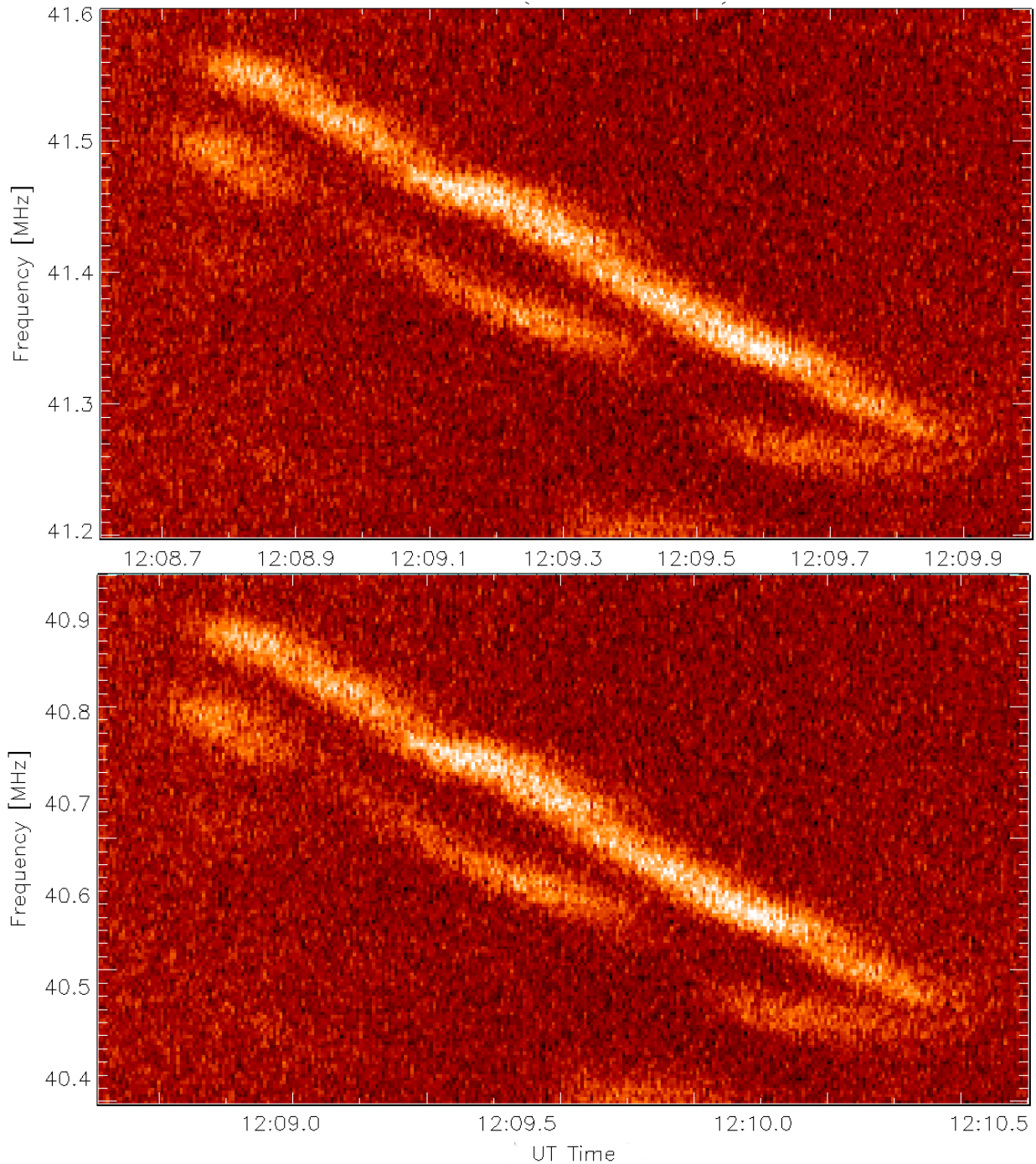

Figure 15. Examples of thin-type stria occurring below 41.5 MHz on 13 April 2013.

In plasma emission, the intensity at a single frequency is related to the duration over which wave growth occurs at the single, corresponding density. Thus, extending the homogeneous region over which the plasma emits strengthens the signal. Considering a single stria, if its exciter were to encounter regions of varying density the intensity of emission would vary accordingly. We consider an exciter that maintains a constant velocity of propagation, generally moving to regions of lower density as it moves out from the Sun. As it enters a more extended homogeneous region the intensity would increase while the spectral drift would decrease. Similarly, when the exciter exited that region into one of abruptly lower density the emission intensity would decrease and the spectral drift rate would increase. Such variations can be seen in the examples of Figure 15. If the source regions of these thinner striae had lower optical depths as those of the stria with greater bandwidth, the emission would have narrow bandwidth, it would be less intense, and it would be more easily affected by small changes in density, just as is observed here. It is important to note that these thin-type stria groups occur along with the wider-bandwidth striations explored before, at similar frequencies and times. Therefore their source locations must be in similar but slightly different physical conditions.

A successful theory for Type-III burst radio emission that would seem to lend itself to fine structures such as striae is "stochastic growth theory" (SGT; *e.g.* Cairns and Robinson, 1998), in which it is assumed that the growth rate of Langmuir waves is a stochastic variable because the energy source is in a state of marginal stability. SGT predicts that Langmuir waves should be intrinsically bursty, as observed in interplanetary Type-III bursts in the solar wind. Li, Cairns, and Robinson (2011a; 2011b; 2012) have modelled fine structure in Type-III bursts and shown that they can reproduce striae-like features from enhanced density structures along the beam path and either electron- or ion-temperature enhancements, but in all cases they find that the second harmonic emission should be stronger than emission at the fundamental of the plasma frequency, and this is inconsistent with the observations. The high degrees of circular polarization shown by the striae in our observations strongly suggest fundamental emission, and we have LWA observations of isolated fundamental–harmonic pairs in which striae only occur in the fundamental band. This result will be discussed further in a future study.

## 4. Summary and Conclusions

The solar data obtained with the LWA1 station are of superb resolution and signal-to-noise ratio. This high-quality data can be used to fully explore the dynamics of solar radio transients and to test existing theories of their emission. As the LWA continues to grow, the advent of imaging concurrent with spectral observations will increase the value of this instrument for solar work. We now take the opportunity to present an analytically derived plausible scenario for the conditions necessary for Type-IIIb storms.

The LWA1 data has been used to add an argument against the theory of Type-IIIb emission from electrons trapped in electrostatic wave sidebands. Takakura and Yousef (1975) argued that stria emission could be instead due to variations in density along the path of the electron beams. Subsequent numerical modeling by Kontar (2001) and Li *et al.* (2012) showed that stria-like fine structure can arise from density inhomogeneities. Other studies were able to produce similar emission patterns from temperature inhomogeneities (Li *et al.*, 2011a, b) or from turbulence (Loi, Cairns, and Li, 2014) along the exciter's path. Che, Goldstein, and Viñas (2014) showed, through particle-in-cell simulations, how two-stream instabilities such as that driven by traveling electron beams can produce long-lived turbulence. As mentioned before, several workers report that a fraction of Type-IIIb bursts can be associated with regular, smooth Type-III emission in a fundamental–harmonic relation. Although the harmonic relationship is inverted in the aforementioned simulations, it is clear that electron beams producing Type-IIIb bursts must travel along field lines which sustain density inhomogeneities and/or increased turbulence on the order or many hours to a few days. For events studied here, we must also account for the sudden onset of the Type-IIIb storm amidst an already occurring Type-III storm despite there being no major event that could be an immediate precursor, changing the conditions along the electron beam paths in an enduring way. We must also address what makes the particular source AR produce Type-IIIb bursts. We propose that the reason that the large CME on 11 April produced an SEP storm is the same reason this AR also produced a Type-III storm: the magnetic field configuration is such that it drives high levels of continued magnetic reconnection. Sustained magnetic reconnection is one way to produce the populations of the seed particles now viewed as a prerequisite for efficient SEP production through diffusive shock acceleration (suprathermals; Laming et al., 2013). A prolonged episode of reconnection would also be an efficient accelerator of electron beams out into the corona, the very drivers of Type-III emission. If density and/or temperature inhomogeneities trigger Type-IIIb emission in the fundamental, and Type-III emission in the harmonic then this particular active region must have certain field lines along which sufficiently turbulent conditions were developed and sustained. It is beyond the scope of this data set to determine a mechanism for such selective field-line enhancements. Perhaps the CME left ample turbulence in its wake, in particular along field lines last connected to the current sheet behind the CME. The temporal delay in the Type-IIIb storm from the large CME occurrence can be due to the directionality of fundamental plasma emission. While inhomogeneities in the corona offer some directional smearing through dispersion, the emission of the fundamental Type-IIIb bursts from electron beams travelling along a particular set of open field lines is more directional that of Type-III bursts (fundamental emission is dipolar in nature, and harmonic emission is quadrupolar). Because these are ground-based observations, the observed Type-IIIb emission would be most favorably observed when solar rotation brings the field lines into the path of Earth. This would be what happened on 14 April, when sporadic Type-IIIb emission became a storm.

This scenario needs but one observable to cement these ideas, that of a statistical correlation of increased Type-III/IIIb activity from AR which produce SEP storms. Further work is then needed to investigate both the large-scale effects, such as Type-III/IIIb occurrence, and the details of radio emission from these bursts, guided by detailed observational constraints provided by the LWA1. We are currently running an observing campaign with the LWA1 to address such questions, as well as further developing automated analysis techniques for such large amounts of data. In addition, we are developing code to process complementary data from the DAM to remove instrumental effects and RFI. In this manner we hope to approach a true understanding of the effects of mildly relativistic electron beams and their implications for other drivers of space weather.

**Acknowledgements**: S. Tun Beltran would like to thank A. Vourlidas and M. Laming for critical discussions of the work, and further express gratitude to the National Academies of Science. The bulk of this work was carried out while he was a National Research Council Fellow at the Naval Research Laboratory. The work was finalized under NRL's Karles Fellowship. Many thanks as well to the LWA1 team, who greatly facilitated the hosting and processing of data for this work. Construction of LWA1 was supported by the Office of Naval Research under Contract N00014-07-C- 0147. Support for operations and continuing development of LWA1 is provided by the National Science Foundation under grants AST-1139963 and AST-1139974 of the University Radio Observatories program.

**Disclosure of Potential Conflicts of Interest**: The authors declare that they have no conflicts of interest.